\documentstyle[psfig]{elsart}
\def\gray{{$\gamma$-ray}}
\def\grays{{$\gamma$-rays}}
\def\etal{{\it et al.}}

\begin{document} 

\begin{frontmatter}

\title{ New Tests of Lorentz Invariance Following from 
Observations of the Highest Energy Cosmic Gamma Rays}
\author{F.W. Stecker}
\address{NASA Goddard Space Flight Center, Greenbelt, MD 20771}
\author{Sheldon L. Glashow}
\address{Boston University, Boston, MA 02215}
%\maketitle
\begin{abstract}               
  
We  use the recent reanalysis of  multi-TeV \gray\ observations of 
Mrk 501 to constrain  the Lorentz invariance breaking
parameter involving the maximum electron velocity. Our  limit is two orders
of magnitude better than that  obtained from the maximum observed cosmic
ray electron energy.

\end{abstract}


\end{frontmatter}

\section{Introduction}

It is occasionally suggested
 that Lorentz invariance (LI) 
may be  only  an
approximate symmetry of nature \cite{sa72}\cite{am98}. A simple 
and self-consistent framework for
analyzing possible departures from exact  LI  was suggested by Coleman and
Glashow\cite{co99}, who assume LI to be broken
perturbatively in the context of conventional quantum field theory. Small
Lorentz noninvariant terms are introduced that are both renormalizable ({\it
i.e.,} of dimension no greater than four) and  gauge invariant under
$SU(3)\times SU(2)\times U(1)$. It is further assumed that the Lagrangian is
rotationally invariant in a preferred frame
which is presumed to be  the rest frame of the
cosmic microwave background.

 Consequent observable manifestations of LI
breaking are either  CPT even or odd, but the former effects are dominant at
high energies. These can be described quite simply in terms of different
maximal attainable velocities   of different particle species as
measured in the preferred frame.  
Indeed, this type of LI breaking within the hadron sector
is one way to circumvent
 the predicted but unseen `GZK cutoff' in the ultrahigh energy 
cosmic-ray spectrum owing to photomeson interactions with 2.7K cosmic
background photons\cite{gzk}
which is expected to produce an effective absorption mean-free-path
for ultrahigh energy cosmic rays in intergalactic space of $< 100$ Mpc
\cite{st68} in the absence of LI breaking.

In this paper, we focus on possible departures from LI in the context of
quantum electrodynamics, whose  effects
conceivably could make the universe transparent to
ultra-high energy  $\gamma$-rays\cite{ki99}.

\section{The LI Breaking Parameter}

We follow the well-defined formalism for 
LI breaking discussed  in reference\cite{co99}.  Within this scenario, 
the maximum attainable velocity of an
electron  need not equal the {\it in vacua\/} velocity of light,
{\it i.e.,}  $c_e \ne c_\gamma$. The physical
consequences of this violation of LI depend on the sign of the difference. 
We define

\begin{equation}
c_{e} \equiv c_{\gamma}(1 +  \delta) ~ , ~ ~~~0< |\delta| \ll 1\;,  
\end{equation}

\noindent
and consider the two cases of positive and negative values of $\delta$
separately. 

{\it Case I:} If $c_e<c_\gamma$ ($\delta \le 0$), 
the decay of a photon into an electron-positron pair is kinematically allowed
for photons with energies exceeding

\begin{equation}
E_{\rm max}= m_e\,\sqrt{2/|\delta|}\;. 
\end{equation}

\noindent
The  decay would take place rapidly, so that photons with energies exceeding 
$E_{\rm max}$ could not be observed either in the laboratory or as cosmic 
rays. From the fact that photons have been observed with energies   
$E_{\gamma} \ge$ 50~TeV from the Crab nebula\cite{ta98}, we deduce for this 
case that $E_{\rm max}\ge 50\;$TeV, or that -$\delta < 2\times  
10^{-16}$.

{\it Case II:}  Here we are concerned with the remaining possibility, where
 $c_e>c_\gamma$ ($\delta \ge 0$) and electrons become superluminal if their
energies exceed $E_{\rm max}/2$.
Electrons traveling faster than light will emit light  at all frequencies by a
process of `vacuum \v Cerenkov radiation.' This process occurs rapidly, so
that superluminal electron energies
 quickly approach $E_{\rm max}/2$. 
 However, because electrons have been seen in the cosmic radiation 
with energies 
up to $\sim\,$1~TeV\cite{ni80}, it follows that 
 $E_{\rm max} \ge 2$~TeV, which leads to 
 an upper limit on $\delta$ 
for this case of $1.3 \times 10^{-13}$. We note that this limit
is  three orders of
magnitude weaker than the limit obtained for  Case I. 

In this note, we show how  stronger bounds on $\delta$ can be set
through searches  for  energetic cosmic ray photons. For case I, the
discussion is trivial: The mere detection of cosmic $\gamma$-rays
   with energies
greater that 50~TeV from sources within our galaxy would improve the bound on
$\delta$. The situation for case II is more interesting.

If LI is broken so that $c_e>c_\gamma$, the threshold energy for the pair 
production process $\gamma + \gamma \rightarrow e^+ + e^-$ is altered 
because the square of the four-momentum becomes

\begin{equation}
2\epsilon E_{\gamma}(1 - \cos \theta) - 2E_{\gamma}^2\delta = 4\gamma^2m_{e}^2 >4 m_{e}^2
\end{equation}

\noindent where $\epsilon$ is the energy of the low energy (infrared) photon and $\theta$
is the angle between the two photons. The second term on the left-hand-side
comes from the fact that $c_{\gamma} =  
\partial E_{\gamma}/\partial p_{\gamma}$.

For head-on collisions ($\cos \theta = -1$) the minimum low energy photon
energy for pair production becomes 

\begin{equation}
\epsilon_{min} = m_{e}^2/E_{\gamma} +  (E_{\gamma}\,\delta)/2
\end{equation}

\noindent It follows that the condition for a significant 
increase in the energy
threshold for pair production is $E_\gamma\ge E_{\rm max}$, or equivalently, 

\begin{equation}
\delta \ge 2m_e^2/E_{\gamma}^2\,.
\end{equation}

\section{Recent Observations of the Blazar Mrk 501}

The highest energy extragalactic \gray\ sources in the known universe are
the active galaxies called `blazars,' objects that emit jets of relativistic 
plasma aimed directly at us with typical bulk Lorentz factors $\sim\,$10.
Those blazars  known as X-ray selected BL Lac objects (XBLs), or
alternatively as high frequency BL Lac objects (HBLs), are expected to emit
photons in the multi-TeV energy range\cite{st96}, but  only the nearest ones
are expected to be observable, the others being hidden by intergalactic
absorption\cite{st92}.

 Cosmic photons with the highest energies yet observed originated in a 
powerful flare coming from the object known as Markarian (Mrk) 
501\cite{ah99}. Its  spectrum  was interpreted by 
Konopelko \etal\cite{ko99} as most naturally showing the 
absorption effect predicted using the calculations of Stecker and
De Jager\cite{st98}, based on the infrared spectra predicted by Malkan and
Stecker\cite{ms98}. This absorption is the result of electron-positron 
pair production by interactions of the multi-TeV \grays\ from Mrk 501 with
intergalactic infrared photons.

An analysis of direct far infrared data from the {\it COBE-DIRBE}
satellite\cite{ha98} was alleged to imply that there should be more
absorbtion than evidenced in the
 Mrk 501 spectrum. Indeed,
 LI breaking was invoked\cite{pr00} as one of various remedies for this
 supposed conflict. 
However, newer work on the infrared background\cite{ms01} and a reanalysis of 
the Mrk 501 data with better energy resolution\cite{ah01} indicate that the
Mrk 501 spectrum is consistent with what one would expect from intergalactic
absorption\cite{sd01}. 

 Intrinsic absorption within Mrk 501  is apt to be 
negligible because it is a giant elliptical galaxy with little dust to emit 
 infrared radiation and because BL Lac objects have little gas (and 
therefore most likely little dust) in their nuclear regions. It also appears
that \gray\ emission in blazars takes place at superluminal knots in 
the jet downstream of the core and at any putative accretion disks\cite{jo01}. 
Thus, it appears that the Mrk 501 \gray\ spectrum above $\sim$ 10 TeV can 
be understood as a result  of intergalactic absorption. 
We therefore interpret the Mrk 501 data as evidence for intergalactic
absorption with no indication of LI breaking up to a photon energy of 
$\sim\,$20~TeV.

\section{Conclusion}

If, as we argue above,
 there is no significant decrease in the optical depth to Mrk 501 for 
$E_{\gamma} \le 20$~TeV,
then it follows from eq. (5) that $\delta \le 2(m_{e}/E_{\gamma})^2 = 
1.3 \times 10^{-15}$. This constraint  is two orders of 
magnitude stronger  than that obtained from 
cosmic-ray electron data as discussed
in section II for the case when $\delta \ge 0$ 
(Case II). Our result for
Case I ($\delta \le 0$) is $|\delta| \le 2\times 10^{-16}$.

Further tests of LI could emerge from future observations.
We mentioned earlier that the detection of galactic $\gamma$-rays with energies
greater than 50 TeV would strengthen the bound on $\delta$ for Case I.
As for Case II, the detection of cosmic 
$\gamma$-rays above $100(1+z_{s})^{-2}$ TeV from an extragalactic source
at a redshift $z_{s}$,  would be strong evidence for LI breaking with
$\delta \ge 0$. 
This is because the very large density ($\sim\,$400~cm$^{-3}$) of 3K cosmic 
microwave photons would otherwise absorb \grays\ of energy $\ge 100$ TeV 
within a distance of $\sim$ 10 kpc,  with  this critical energy  
reduced by a factor of $\sim (1+z_{s})^2$ for sources at redshift 
$z_{s}$\cite{st69}.

\end{document}